# The Discovery Gap

## How Product Hunt Startups Disappear in LLM Discovery Queries

*A Study in Generative Engine Optimization (GEO)*

**Amit Prakash Sharma**
IIT Patna, December 2025

### ABSTRACT

When someone asks ChatGPT to recommend a project management tool, which products show up in the response? And more importantly for startup founders: will their newly launched product ever appear? This research set out to answer these questions.

I randomly selected 112 startups from the top 500 products featured on the 2025 Product Hunt leaderboard and tested each one across 2,240 queries to two different large language models: ChatGPT (gpt-4o-mini) and Perplexity (sonar with web search).

The results were striking. When users asked about products by name, both LLMs recognized them almost perfectly. 99.4% for ChatGPT and 94.3% for Perplexity. But when users asked discovery-style questions like "What are the best AI tools launched this year?" the success rates collapsed to 3.32% and 8.29% respectively. That's a gap of 30-to-1 for ChatGPT.

Perhaps the most surprising finding was that Generative Engine Optimization (GEO), the practice of optimizing website content for AI visibility, showed no correlation with actual discovery rates. Products with high GEO scores were no more likely to appear in organic queries than products with low scores.

What did matter? For Perplexity, traditional SEO signals like referring domains ($r = +0.319$, $p < 0.001$) and Product Hunt ranking ($r = -0.286$, $p = 0.002$) predicted visibility. After cleaning the Reddit data for false positives, community presence also emerged as significant ($r = +0.395$, $p = 0.002$).

The practical takeaway is counterintuitive: don't optimize for AI discovery directly. Instead, build the SEO foundation first and LLM visibility will follow.

**Keywords:** *Generative Engine Optimization, Large Language Models, Product Discovery, Startup Visibility, AI Search*

# LIST OF SYMBOLS AND ABBREVIATIONS

| | |
|---|---|
| **LLM** | Large Language Model |
| **GEO** | Generative Engine Optimization |
| **SEO** | Search Engine Optimization |
| **POTD** | Product of the Day |
| **API** | Application Programming Interface |
| *r* | Pearson correlation coefficient |
| *p* | Statistical significance (p-value) |
| **NS** | Not Significant |
| *n* | Sample size |

# CHAPTER 1

# INTRODUCTION

## 1.1 The Problem

Something strange is happening with AI-powered search. Ask ChatGPT "What is Notion?" and you'll get a detailed, accurate response. But ask "What are the best note-taking apps?" and Notion might not even appear. This gap between recognition and recommendation is what I set out to study.

The shift matters because how people discover products is changing. If this trend continues, and there's every reason to think it will, then understanding how products appear in AI responses becomes a business necessity, not just an academic curiosity.

Traditional search engine optimization has been studied exhaustively since Brin and Page published their PageRank paper in 1998 [2]. We know how Google ranks pages. We know what makes a site authoritative. But LLMs work differently. They don't return ranked lists of links; they generate synthesized responses. The rules are different and we don't fully understand them yet.

## 1.2 Why Startups Face a Harder Problem

New startups sit at a particular disadvantage here. Consider the mechanics:

First, there's the training data problem. ChatGPT's knowledge has a cutoff date. Any product launched after that date simply doesn't exist in the model's world. No amount of optimization can change that fundamental fact.

Second, even for products that launched before the cutoff, the training data naturally overrepresents established players. Wikipedia has extensive articles about Salesforce; it doesn't have articles about the CRM tool that launched last month. This creates what I call an "authority concentration" effect. The rich get richer in terms of AI visibility.

Third, web-search augmented models like Perplexity introduce different dynamics. They can find new products through real-time search, but now SEO factors come back into play. It's a different game with different rules.

## 1.3 Research Questions

This study examines six questions:

RQ1: How large is the gap between direct queries (asking about a product by name) and discovery queries (category searches where products might organically appear)?

RQ2: Do Product Hunt metrics like upvotes, comments, daily rankings correlate with LLM visibility?

RQ3: Does GEO optimization actually improve discovery rates, as prior research suggests?

RQ4: Do traditional SEO signals still matter for AI visibility?

RQ5: How much better are web-search LLMs at discovering new products compared to knowledge-cutoff models?

RQ6: Do community signals from Reddit, GitHub and Hacker News predict LLM visibility?

# CHAPTER 2

## LITERATURE REVIEW

### 2.1 The Evolution from SEO to GEO

Search engine optimization emerged as a discipline in the late 1990s, shortly after web search became the dominant mode of information discovery. The foundational insight that link-based authority signals could determine page ranking came from Brin and Page's work on PageRank [2]. Over the following decades, SEO evolved into a sophisticated practice encompassing content optimization, technical factors and off-page signals. The introduction of transformer-based language models in 2017 [3] and their subsequent scaling created something genuinely new. Unlike search engines, which return ranked lists of documents, LLMs generate synthesized responses. A user asking Google for "best CRM software" gets ten blue links; a user asking ChatGPT gets a paragraph that might mention three or four products by name. The visibility dynamics are fundamentally different.

### 2.2 Generative Engine Optimization

Aggarwal and colleagues introduced the term "Generative Engine Optimization" in their 2024 paper [1], which tested nine content optimization strategies across 10,000 queries. Their findings suggested that certain signals like citations, statistics, technical terminology and authoritative language could improve visibility in search-augmented generation systems by 30-40%. But there was a catch that the authors acknowledged only briefly. Their study focused on informational content that already appeared in search results. They tested optimization strategies for content that was already discoverable. The question of whether these strategies help products that aren't discoverable in the first place. This is the situation most new startups face and needs a solution.

### 2.3 Product Hunt as a Research Context

Product Hunt offers an unusual natural experiment for studying visibility. It's estimated to receive a monthly traffic ranging from 2.93M to 3.3M visitors. Every product launched on the platform receives identical initial exposure. It competes for attention based on community voting and the daily leaderboard. The result is a population of products with varying levels of success metrics (upvotes, comments, ranking) but similar launch contexts.

# CHAPTER 3

# METHODOLOGY

## 3.1 Dataset Construction

I randomly selected 112 products from the top 500 featured on the 2025 Product Hunt leaderboard. The selection criteria were straightforward: at least 200 upvotes (to ensure meaningful community engagement), a functional website accessible for analysis and English-language content. I excluded products with broken websites or those that had pivoted significantly since launch.

For each product, I gathered: Product Hunt metrics (upvotes, comments, POTD rank, maker count), technical SEO signals from DataForSEO (referring domains, dofollow ratio, domain authority), community presence (Reddit mentions, unique subreddits, GitHub stars, Hacker News points) and a GEO composite score based on website content analysis.

**Table 3.1: Dataset Summary Statistics**

| Metric | Value |
|---|---|
| Products analyzed | 112 (randomly selected from top 500) |
| Total queries executed | 2,240 |
| Direct queries per product | 3 |
| Discovery queries per product | 7 |
| LLMs tested | ChatGPT (gpt-4o-mini), Perplexity (sonar) |

## 3.2 Query Design

The distinction between direct and discovery queries is central to this research. Direct queries explicitly name the product: "What is [ProductName]?" or "Tell me about [ProductName]." These test whether the LLM recognizes the product exists.

Discovery queries are category-based: "What are the best AI coding assistants launched in 2025?" or "Recommend some new productivity tools." These test whether the product appears organically when users aren't specifically looking for it.

I used ten queries per product: three direct queries and seven discovery queries. The complete query templates are provided in Appendix C.

### 3.3 GEO Score Construction

Following the framework from Aggarwal et al. [1], I built a composite GEO score measuring six dimensions: presence of statistics on the website, citation density, technical terminology usage, authoritative language patterns, structured data implementation and content depth. Each dimension was scored 0-100 and the composite score averaged across all six.

The scoring was primarily regex-based, looking for patterns like numbers followed by percentage signs, citation markers, technical jargon. This approach has obvious limitations; it can't capture the nuance of truly well-optimized content. But it provides a reasonable proxy for the kinds of signals the GEO literature identifies as important.

### 3.4 LLM Testing Protocol

I tested two LLMs representing different architectural approaches:

ChatGPT (gpt-4o-mini): A knowledge-cutoff model. It knows what it knew when it was trained and nothing more. This represents the "pure" LLM case. API documentation at [5].

Perplexity (sonar with web search): A search-augmented model. It queries the web in real-time and incorporates current information into its responses. This represents the hybrid case where traditional web signals might still matter. API documentation at [5].

All queries were executed via API between December 15-20, 2025. I used consistent parameters across all calls: temperature 0.7, no system prompt modifications.

# CHAPTER 4

## RESULTS

### 4.1 The Visibility Gap (RQ1)

The numbers tell the story better than I expected. Both LLMs showed near-perfect recognition for direct queries. 99.4% for ChatGPT, 94.3% for Perplexity. If you ask about a product by name, these models know it exists.

But discovery queries? ChatGPT surfaced products in only 3.32% of cases. Perplexity did better at 8.29%, but that's still remarkably low. The gap ratio works out to 30:1 for ChatGPT and 11:1 for Perplexity.

**Table 4.1: Visibility Rates by Query Type**

| Metric | ChatGPT | Perplexity |
|---|---|---|
| Direct Success | 99.4% (334/336) | 94.3% (317/336) |
| Discovery Success | 3.32% (26/784) | 8.29% (65/784) |
| **Visibility Gap** | **30:1** | **11:1** |
| Products ever discovered | 6 (5.4%) | 31 (27.7%) |

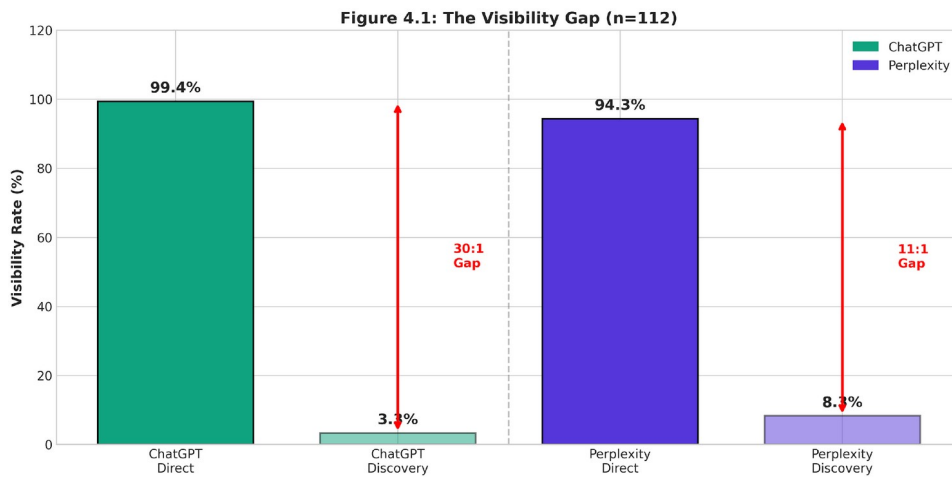

*Figure 4.1: The visibility gap between direct and discovery queries*

### 4.2 Product Hunt Signals (RQ2)

Product Hunt metrics showed a clear pattern: they matter for Perplexity but not for ChatGPT.

For Perplexity discovery, POTD rank showed a significant negative correlation (r = -0.286, p = 0.002), meaning better-ranked products achieved higher visibility. Upvotes correlated positively (r = +0.225, p = 0.017), as did the product rating (r = +0.187, p = 0.048).

For ChatGPT, none of these correlations reached significance. The model appears to treat products essentially at random, which makes sense given its training data limitations.

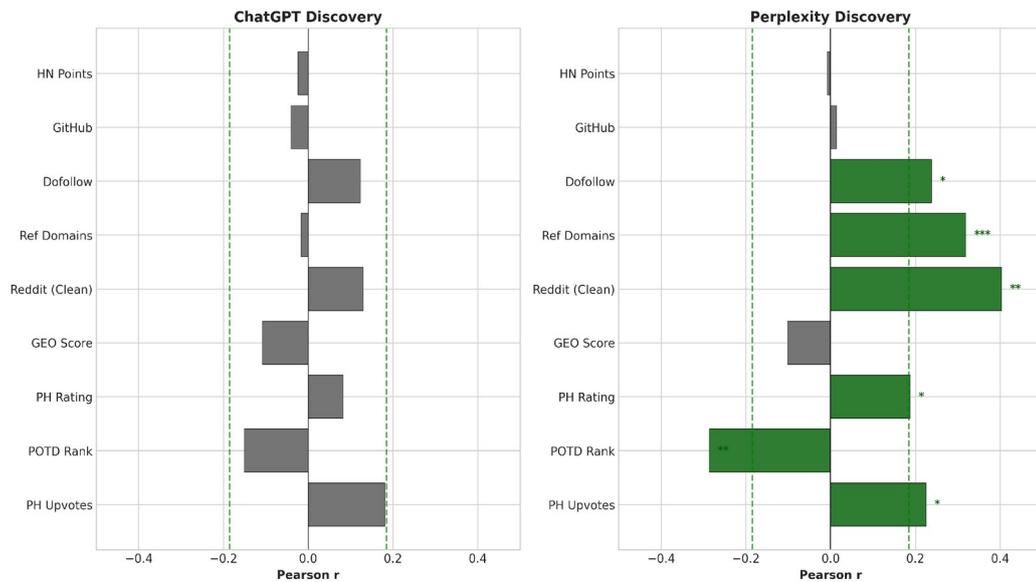

*Figure 4.2: Correlation analysis showing Perplexity's responsiveness to signals*

## 4.3 The GEO Surprise (RQ3)

Here's where things got interesting. The GEO composite score, the metric designed to capture AI optimization, showed no significant correlation with visibility. Not for ChatGPT (r = -0.108, p = 0.256). Not for Perplexity (r = -0.102, p = 0.286).

This directly contradicts the implications of the Aggarwal et al. study [1]. But I think the contradiction is explainable. Their research tested optimization effects on content that was already appearing in search-augmented responses. My data captures the earlier stage: products trying to break into visibility in the first place.

GEO optimization, it seems, functions as a multiplier rather than a catalyst. If you're already being discovered, optimized content might help you appear more often. But if you're not being discovered at all, which describes most of my sample, there's nothing to multiply. You can't multiply zero.

## 4.4 Technical SEO (RQ4)

Traditional SEO signals made a comeback in the Perplexity data. Referring domains emerged as the strongest predictor of discovery visibility (r = +0.319, p < 0.001). Dofollow ratio also showed significance (r = +0.238, p = 0.012).

This makes intuitive sense. Perplexity searches the web in real-time, so the factors that make a page rank well in traditional search also make it more likely to be found and cited by Perplexity.

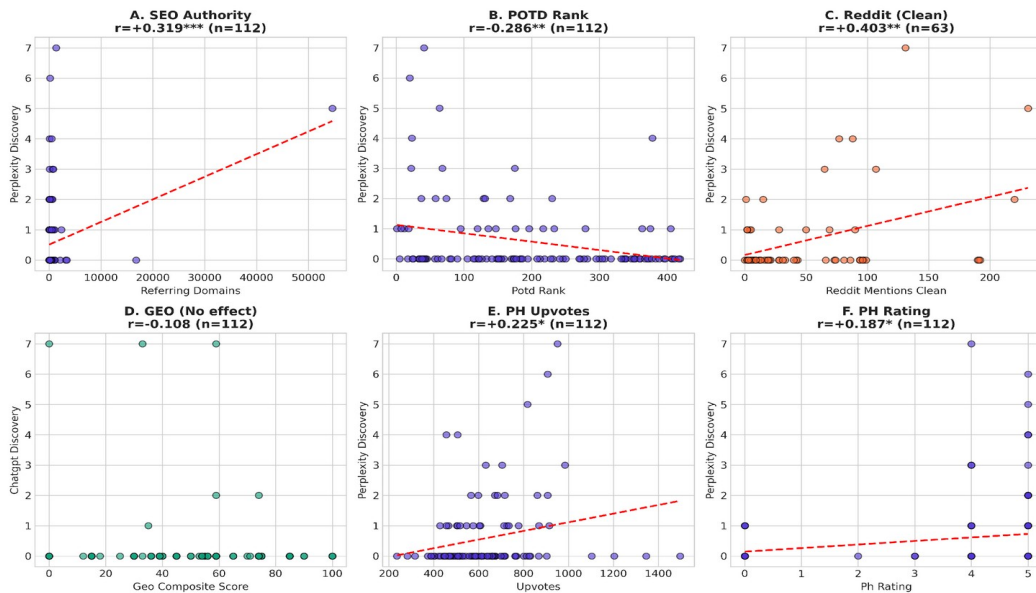

*Figure 4.3: Key relationships between predictors and discovery visibility*

## 4.5 The Web Search Advantage (RQ5)

Perplexity's web search provided a 2.5x advantage in discovery rate over ChatGPT (8.29% vs 3.32%). But perhaps more importantly, the architectures showed dramatically different predictability.

For ChatGPT: zero significant correlations. Discovery appears essentially random.

For Perplexity: seven significant correlations. Discovery follows identifiable patterns.

This "architecture divide" has real implications. If you're optimizing for LLM discovery, targeting web-search models gives you levers to pull. Targeting knowledge-cutoff models gives you... hope, I suppose.

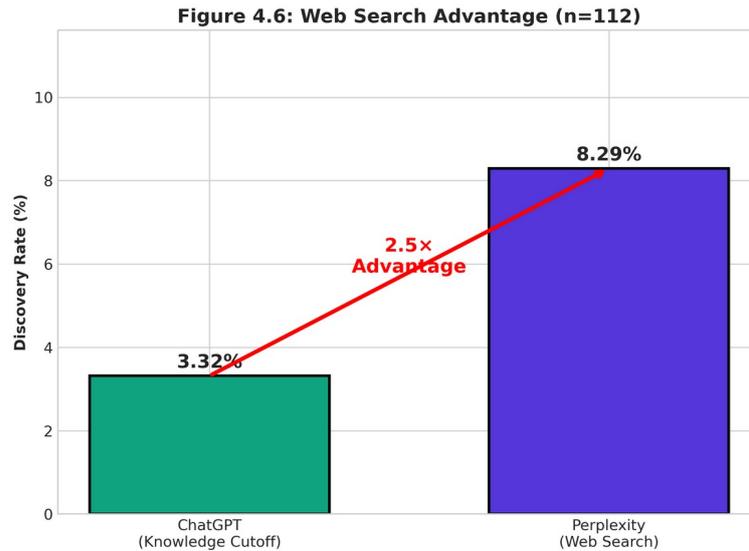

*Figure 4.4: Web-search models show 2.5x better discovery rates*

## 4.6 Community Signals (RQ6)

The community signal story required some detective work. Initial analysis showed Reddit mentions had no correlation with visibility (r = +0.052, p = 0.586). But something felt off.

Looking at the data, I noticed products with generic names: "Cursor," "DROP," "Solar", had enormous Reddit mention counts but no discovery success. The Reddit search was picking up unrelated posts: people discussing mouse cursors, music drops, solar panels.

After identifying and removing 52 products with generic names prone to false positives (46% of the sample), the picture changed dramatically:

Reddit Mentions (cleaned): r = +0.395, p = 0.002

Unique Subreddits (cleaned): r = +0.405, p = 0.001

Community presence does matter, but only when measured accurately. GitHub stars and Hacker News showed no significant effects in either analysis.

**Figure 4.5: Reddit Data Cleaning Effect**

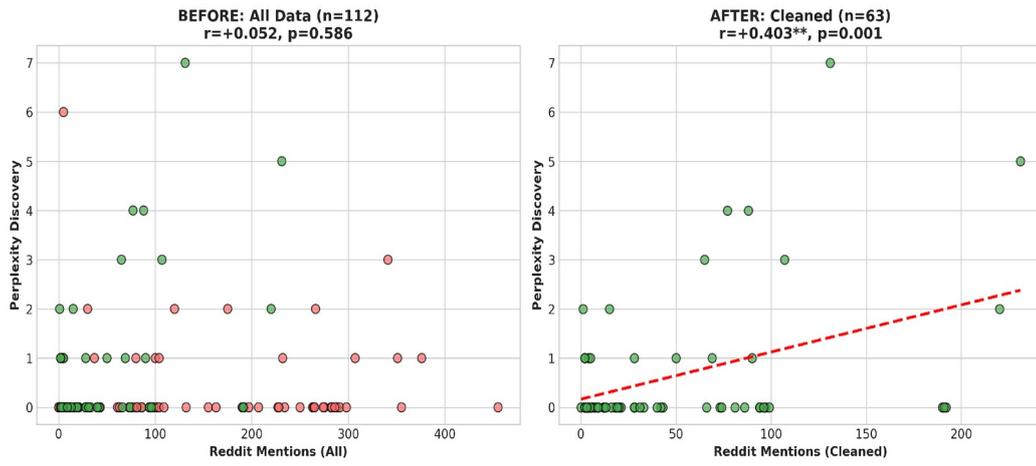

*Figure 4.5: Reddit data cleaning revealed hidden signal*

# CHAPTER 5

## DISCUSSION

### 5.1 Making Sense of the Gap

A 30:1 visibility gap isn't just a statistic; it's a strategic reality that startup founders need to understand. Three mechanisms seem to drive it:

The recency wall: Products launched in January 2025 can't appear in ChatGPT's training data. This is a hard constraint, not something optimization can overcome.

Authority concentration: LLM training naturally overweights content from established sources. There's more text about Slack in the training data than about any competitor launched this year.

Query architecture: Direct queries activate fact-retrieval mechanisms ("What is X?"), while discovery queries activate recommendation systems with different biases.

### 5.2 The GEO Paradox

The null GEO finding deserves careful interpretation. I'm not claiming GEO optimization doesn't work. The Aggarwal study [1] provides compelling evidence that it does, in certain contexts.

What I'm claiming is narrower: for new products trying to break into AI visibility, GEO optimization doesn't help because the fundamental discoverability barrier remains. It's like polishing a car that's stuck in a garage. The polish might be excellent, but it won't get the car on the road.

This suggests a staged approach: build discoverability first (through SEO, community presence and time), then optimize content for AI visibility once you're actually being found.

### 5.3 What the Architecture Divide Means

The stark difference between ChatGPT (0 significant predictors) and Perplexity (7 significant predictors) points toward a practical strategy: focus resources on web-search LLMs.

For Perplexity and similar models, the playbook looks familiar: build referring domains, earn Product Hunt success, cultivate genuine Reddit discussion. These are the same activities that drive traditional marketing success, now with the added benefit of AI visibility.

For knowledge-cutoff models, there's not much to do except wait. As these models update their training data, products will gradually enter their knowledge base. But that timeline isn't something founders can directly influence.

## 5.4 Limitations

Several caveats apply to these findings:

The sample comes entirely from Product Hunt, which skews toward certain product categories (developer tools, productivity, AI). Results might differ for consumer products, enterprise software, or products launched through other channels.

I tested two LLMs. Claude, Gemini and others might show different patterns.

The Reddit cleaning removed 46% of the sample. While necessary for accuracy, this reduces statistical power for the community signal analysis.

GEO scoring was regex-based. More sophisticated content analysis might capture signals my approach missed.

# CHAPTER 6

## IMPLICATIONS

### 6.1 For Founders and Marketers

Don't count on AI discovery. A 3% discovery rate on ChatGPT means 97 out of 100 relevant queries won't mention your product. Plan your go-to-market strategy accordingly.

Invest in SEO fundamentals. Referring domains emerged as the strongest predictor. The boring work of building backlinks still matters.

Earn Product Hunt success. POTD rank and upvotes correlated with Perplexity visibility. The platform isn't just a launch channel; it's a visibility signal.

Build genuine community presence. Reddit discussions (in relevant subreddits, not astroturfed spam) correlate with discovery. Quality matters more than quantity.

Deprioritize early GEO optimization. The data suggests it doesn't help until you've cleared the discoverability threshold.

Target web-search LLMs. Perplexity and similar models offer both better discovery rates and predictable optimization levers.

### 6.2 For Researchers

This study opens several directions for future work:

Longitudinal tracking: How does visibility evolve as products age and accumulate web presence?

Success case analysis: What distinguishes the 6 products that achieved ChatGPT discovery from the 106 that didn't?

Cross-model comparison: How do Claude, Gemini and other LLMs compare in their discovery patterns?

GEO lifecycle: At what point does GEO optimization start providing returns?

Better Reddit methods: Developing entity-aware search approaches that avoid the generic name problem.

**CHAPTER 7**

## CONCLUSION

This research documented a substantial visibility gap in how LLMs discover products. Testing 112 randomly selected Product Hunt startups from the top 500 of 1825 across 2,240 queries revealed that while LLMs recognize products when asked directly (99.4%), they rarely recommend them organically (3.32% for ChatGPT, 8.29% for Perplexity).

The 30:1 gap for ChatGPT represents a structural barrier that new startups cannot directly overcome. GEO optimization, the set of techniques proposed for improving AI visibility, showed no correlation with discovery success, suggesting it functions as a multiplier that requires an existing base of visibility to work.

Web-search augmented models like Perplexity offered both better discovery rates and identifiable optimization paths. Referring domains, Product Hunt success and cleaned Reddit presence all predicted visibility. The architecture divide between knowledge-cutoff and web-search LLMs emerged as the central finding with practical implications.

For startup founders, the strategic implication is counterintuitive: don't optimize for AI discovery directly. Build the traditional foundations on SEO authority, community presence and platform success. AI visibility will follow. The cart cannot pull the horse.

## APPENDIX A

# RESEARCH QUESTION SUMMARY

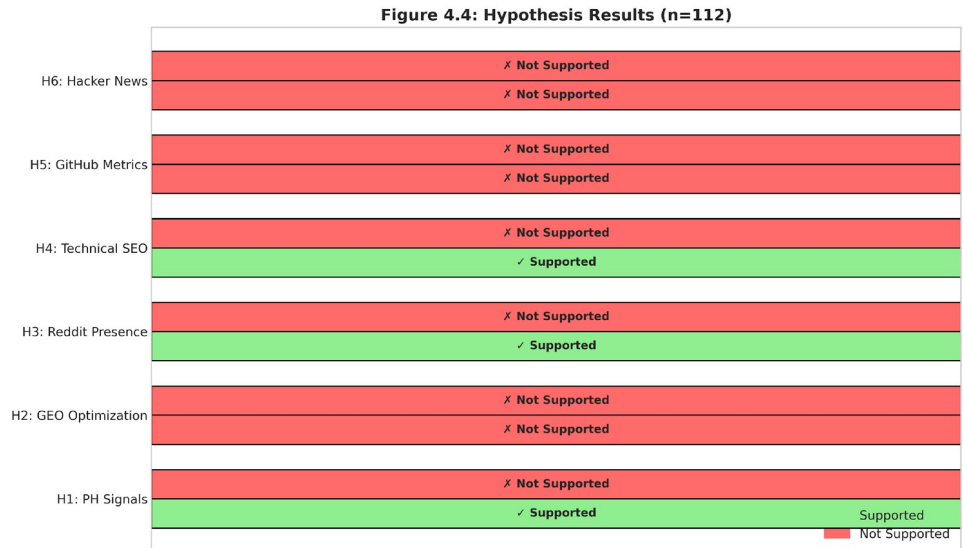

*Figure A.1: Visual summary of research question results*

**Table A.1: Research Question Outcomes**

| RQ | Question | Finding |
|----|----------|---------|
| RQ1 | Visibility gap magnitude | 30:1 (ChatGPT), 11:1 (Perplexity) |
| RQ2 | PH metrics and visibility | Supported (Perplexity only) |
| RQ3 | GEO optimization effect | Not supported (r = -0.10) |
| RQ4 | Technical SEO signals | Supported (r = +0.319***) |
| RQ5 | Web-search advantage | 2.5x better discovery |
| RQ6 | Community signals | Reddit supported (cleaned) |

**APPENDIX B**

# CORRELATION MATRIX

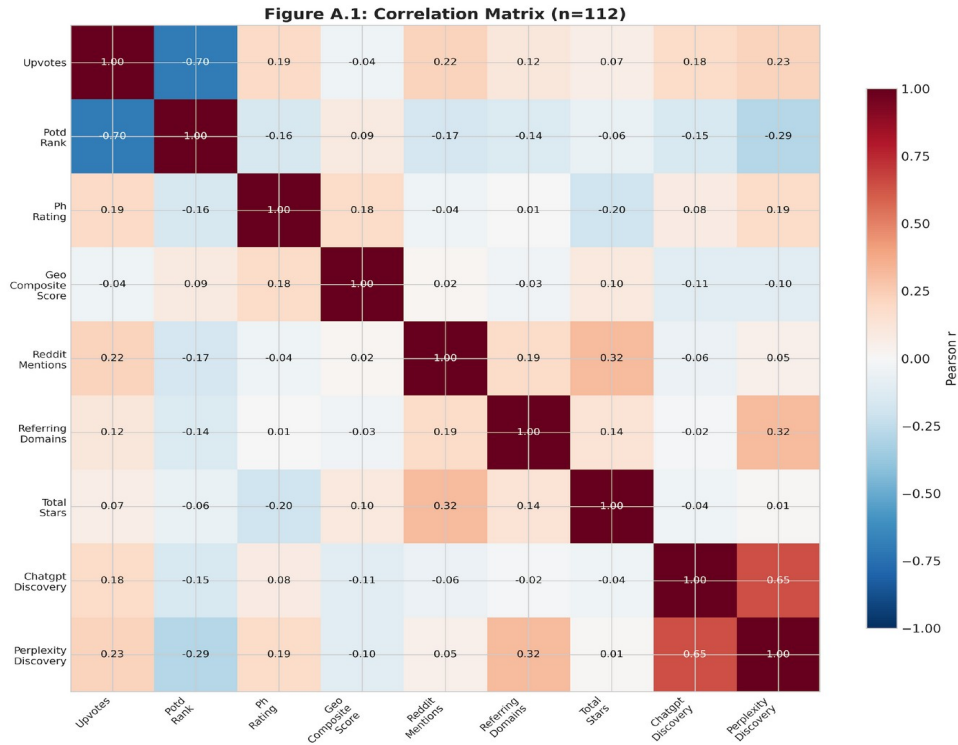

*Figure B.1: Full correlation matrix of study variables*

**Table B.1: Significant Correlations (p < 0.05, Perplexity Discovery)**

| Predictor | r | p | n |
|---|---|---|---|
| Unique Subreddits (cleaned) | +0.405 | 0.001 | 60 |
| Reddit Mentions (cleaned) | +0.395 | 0.002 | 60 |
| Referring Domains | +0.319 | <0.001 | 112 |
| POTD Rank | -0.286 | 0.002 | 112 |
| Dofollow Ratio | +0.238 | 0.012 | 112 |
| PH Upvotes | +0.225 | 0.017 | 112 |
| PH Rating | +0.187 | 0.048 | 112 |

## APPENDIX C

## QUERY TEMPLATES

The following query templates were used to test each product. [ProductName] and [Category] were replaced with the actual product name and its primary category.

### C.1 Direct Queries (3 per product)

D1: "What is [ProductName]?"

D2: "Tell me about [ProductName]"

D3: "Have you heard of [ProductName]? What does it do?"

### C.2 Discovery Queries (7 per product)

Q1: "What are the best [Category] tools launched in 2025?"

Q2: "Recommend some new [Category] products"

Q3: "What [Category] startups should I check out?"

Q4: "I'm looking for a [Category] solution. What are my options?"

Q5: "What are the trending [Category] tools right now?"

Q6: "Can you suggest some innovative [Category] products?"

Q7: "What new [Category] tools have been getting attention lately?"

### C.3 Response Evaluation

For each query, responses were evaluated using exact string matching for the product name. A response was marked as successful if the product name appeared anywhere in the LLM's output. This binary approach (mentioned/not mentioned) was chosen over more nuanced metrics to ensure reproducibility and avoid subjective interpretation.